\documentclass[11pt]{article}
\usepackage{amssymb,amsmath,cite,geometry,graphicx}
\catcode`\@=11

\@addtoreset{equation}{section}
\def\theequation{\arabic{section}.\arabic{equation}}
     
\catcode`\@=11
\def\thesection{\arabic{section}.}

\def\appendix{\setcounter{section}{0}
        \def\thesection{Appendix.}
        \def\theequation{\Alph{section}.\arabic{equation}}}
\def\section{\@startsection{section}{1}{\z@}{3.5ex plus 1ex minus
   .2ex}{2.3ex plus .2ex}{\large\bf}}

\long\def\@makefntext#1{\parindent 0cm\noindent
\hbox to 1em{\hss$^{\@thefnmark}$}#1}

\newcommand{\captionfonts}{\small}
\makeatletter  
\long\def\@makecaption#1#2{%
  \vskip\abovecaptionskip
  \sbox\@tempboxa{{\captionfonts #1: #2}}%
  \ifdim \wd\@tempboxa >\hsize
    {\captionfonts #1: #2\par}
  \else
    \hbox to\hsize{\hfil\box\@tempboxa\hfil}%
  \fi
  \vskip\belowcaptionskip}
\makeatother   

\begin{document}
\begin{titlepage}
\vspace{.5in}
\begin{flushright}
 
May 2015\\  
\end{flushright}
\vspace{.5in}
\begin{center}
{\Large\bf
 Can gravitational microlensing\\[.6ex] by vacuum fluctuations be observed?}\\  
\vspace{.4in}
{S.~C{\sc arlip}\footnote{\it email: carlip@physics.ucdavis.edu}\\
       {\small\it Department of Physics}\\
       {\small\it University of California}\\
       {\small\it Davis, CA 95616}\\{\small\it USA}}
\end{center}

\vspace{.5in}
\begin{center}
{\large\bf Abstract}
\end{center}
\begin{center}
\begin{minipage}{5in}
{\small
Although the prospect is more plausible than it might appear, the answer to the title 
question is, unfortunately, ``probably not.''  Quantum fluctuations of 
vacuum energy can focus light, and while the effect is tiny, the distribution 
of fluctuations is highly non-Gaussian, offering hope that relatively rare 
``large'' fluctuations might be observable.  I show that although gravitational 
microlensing by such fluctuations become important at scales much larger than 
the Planck length, the possibility of direct observation remains remote, although 
there is a small chance that cumulative effects over cosmological distances might
be detectable.  The effect is sensitive to the size of the Planck scale, however, and
could offer a new test of TeV-scale gravity.
}
\end{minipage}
\end{center}
\end{titlepage}
\addtocounter{footnote}{-1}

\section{Introduction} 

Quantum gravitational fluctuations of the vacuum are sometimes divided in two
categories \cite{Wu}.  ``Active'' fluctuations are fluctuations of the spacetime geometry
itself; beyond low orders of perturbation theory, their description presumably 
requires a full quantum theory of gravity.  ``Passive'' fluctuations are fluctuations 
of the matter stress-energy tensor that induce fluctuations in the metric 
through the Einstein field equations.  Their effect can be seen, for example, 
in the focusing or defocusing of a beam of light.  A pencil of light---a 
congruence of null geodesics with an affinely parametrized null normal
$\ell^a$---has a cross-sectional area $A$ whose change is
characterized by the expansion
\begin{align}
\theta = \frac{1}{A}\ell^a\nabla_a A = \frac{d\ }{d\lambda}\ln A  ,
\label{a1}
\end{align} 
where $\lambda$ is an affine parameter.
The expansion, in turn, is governed by the Raychaudhuri equation 
\cite{Raychaudhuri,HawkingEllis},
\begin{align}
\frac{d\theta}{d\lambda} = \ell^a\nabla_a\theta 
     = -\frac{1}{2}\theta^2 - \sigma_{ab}\sigma^{ab}
    + \omega_{ab}\omega^{ab} - 8\pi GT_{ab}\ell^a\ell^b ,
\label{a2}
\end{align}
where $\sigma_{ab}$ is the shear, $\omega_{ab}$ is the vorticity, and
the stress-energy tensor on the right-hand side encodes the effects of
passive vacuum fluctuations.  In particular, positive fluctuations 
focus, and thus temporary brighten, the beam.

Now, a typical quantum fluctuation of the stress-energy tensor at a length
scale $L$ has a value $GT\sim \ell_p{}^2/L^4$, where $\ell_p$ is the Planck
length.  Naively, such fluctuations should be negligible at scales 
$L\gg \ell_p$.  Their effects on the expansion $\theta$ were first 
considered at lowest order in \cite{Borgman}, neglecting the nonlinear 
term in the Raychaudhuri equation, and the results were indeed
found to be unobservably small.  As Fewster, 
Ford, and Roman have demonstrated \cite{Fewster,Fewsterb}, though, the 
distribution of fluctuations is highly non-Gaussian, with a sharp 
lower bound and an infinite subexponential positive tail.  In such a 
setting, one must be very careful about expectations.  It was shown in 
\cite{Fewster} that the non-Gaussianity greatly increases the 
probability of nucleating large objects such as primordial black
holes and (perhaps) ``Boltzmann brains,'' and in \cite{CMP} it was
argued that one should expect dramatic effects at the Planck scale.
In this paper I will investigate the possibility that such effects
extend to a scale that might be directly observable.  We shall see,
unfortunately, that while fluctuations cause gravitational lensing at
distances much larger than the Planck length, these are still almost
certainly too small and short-lived to observe.

\section{Vacuum fluctuations and ``gambler's ruin''}

The strategy of Fewster et al.\ \cite{Fewster} was to compute vacuum 
correlation functions of products of many stress-energy tensors and to ask 
for the probability distribution for which these correlators were the moments.  
In two spacetime dimensions, recursion relations among correlators 
allow an exact computation, and the resulting probability distribution 
is unique \cite{Fewsterb}.  In four dimensions, the computation is more 
difficult, but one can obtain an approximate distribution, along 
with very strong restrictions on the behavior of the tail \cite{Fewster}.  
\begin{figure}
\centerline{
\includegraphics[width=4in]{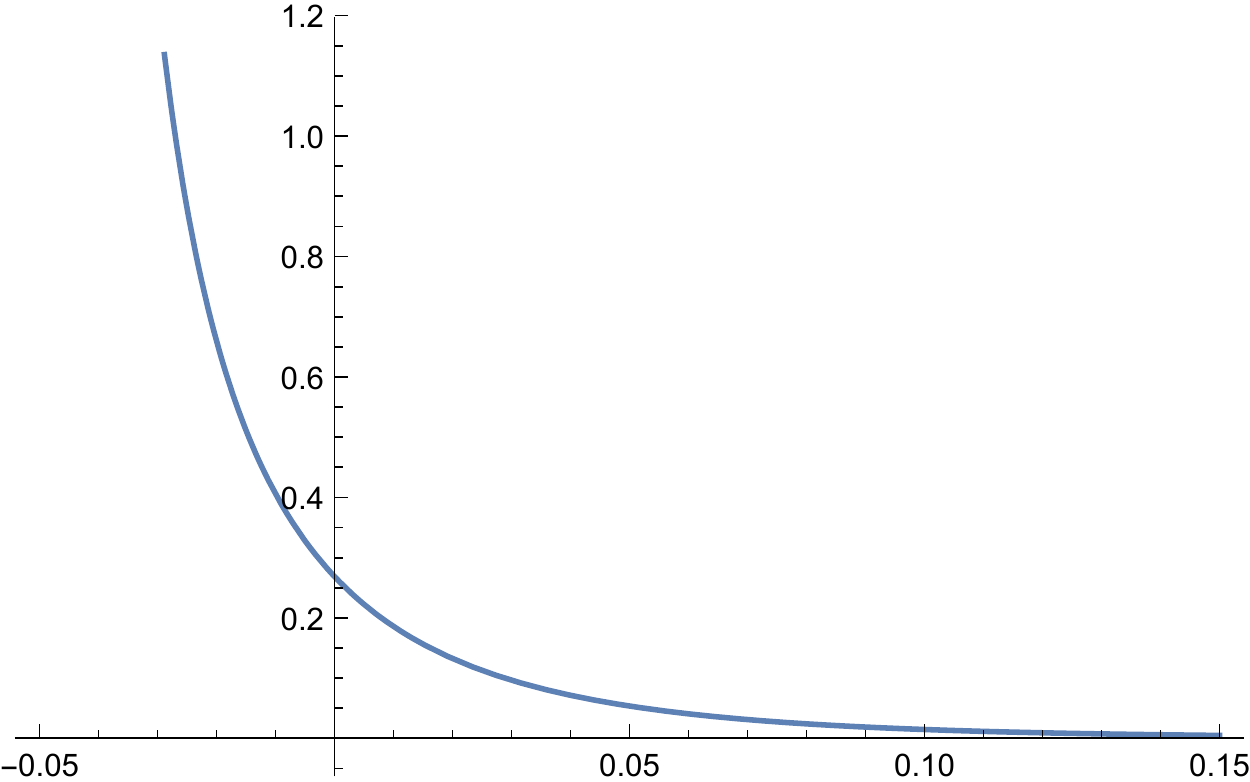}}
\caption{Probability distribution for vacuum fluctuations of electromagnetic energy}
\end{figure}
For vacuum fluctuations of the electromagnetic field, this distribution   
is shown in figure 1, in terms of the dimensionless stress-energy tensor 
at scale $L$,
\begin{align}
x = \frac{16\pi^2}{\hbar} L^4T_{00}  .
\label{b2}
\end{align}
Note that the mass due to a vacuum fluctuations inside a sphere of radius 
$L$ is
\begin{align}
\frac{Gm}{L} \sim \frac{1}{12\pi}\left(\frac{\ell_p}{L}\right)^2\,x   .
\label{bx}
\end{align}
We shall be mainly interested in the long positive tail of the distribution, 
which is well approximated by the equation
\begin{align}
\mathit{Prob}(x) \sim c_0 (x-x_0)^{-2} e^{-a(x-x_0)^{1/3}} ,\qquad
c_0\approx .955,\ a\approx .963,\ x_0\approx -.0472 ,
\label{b1}
\end{align}
where $x_0$ is the (calculated) lower bound on fluctuations.

The full quantum computation of the effects of 
these fluctuations, accounting for all of the nonclassical correlations,
seems extremely difficult (although see \cite{Drago} for a first
effort).  But we can obtain a reasonable approximation by
treating the fluctuations as a classical stochastic process \cite{CMP},
with a random ``kick'' at each time interval $L$  selected from 
the distribution of Fewster et al.

Before proceeding, a few caveats are in order.  First, the results
of \cite{Fewster} were computed for the Minkowski vacuum, and may
fail in regions of high curvature.  We shall be looking at low 
curvature regions, though, so this should not be important.   The
results are also state-dependent; in particular, there are states
in which there is no lower bound on vacuum energy along a null
geodesic \cite{FewsterRoman}, and the argument that follows
will fail for such states.  The length scale $L$ is incorporated by 
smearing the stress-energy tensor over that scale, and the 
details of the distribution may depend on the choice 
of smearing function.  (Fewster et al.\ use a Lorentzian average.)   
Finally---perhaps most seriously for this work---the relevant component
of the stress-energy tensor in the Raychaudhuri equation (\ref{a2})
is a null-null component, while Fewster et al.\ have only computed
distributions for the time-time component.  I will assume that the
behaviors are at least qualitatively similar, as they are in two dimensions 
\cite{Fewsterb}; given the tracelessness of the electromagnetic
stress-energy tensor, one can argue that this must be true by
symmetry considerations, but a more careful analysis would be
useful.\footnote{A further caveat is that the moments of the
probability distribution may not determine the distribution uniquely.
Here, though, we will be dealing almost entirely with the positive
tail, which \emph{is} nearly unique \cite{Fewster}.}

Keeping in mind these caveats, we may proceed as follows
\cite{CMP}.  Let us assume that the initial vorticity of our pencil 
of light is zero; it will then remain zero during propagation 
\cite{HawkingEllis}.  Then the only possible positive term on the
right-hand side of the Raychaudhuri equation---the only term that
can cause a defocusing---comes from the fact that vacuum
fluctuations of $T_{ab}$ can be negative.  But such defocusing 
fluctuations have a lower bound $x_0$, and if $\theta$ is
negative enough, they cannot compete with the focusing 
caused by the nonlinear term $-\theta^2$.  Hence if $\theta$ 
ever becomes sufficiently negative, it will necessarily be driven
to $-\infty$.

This is essentially the phenomenon known in probability 
theory as ``gambler's ruin'' \cite{Coolidge}.  The term has several 
meanings, but here the relevant statement is that if a player with a 
finite amount of money continues to bet against a banker with infinite 
resources, then no matter how favorable the odds, the player will eventually
lose everything.  In our context, the odds for defocusing are favorable%
---most vacuum fluctuations are negative.  But the maximum
amount of defocusing is limited, while focusing is not, so when a 
large enough positive fluctuation eventually occurs, the process 
becomes irreversible.

We may estimate of this effect as follows.  At a fixed scale $L$, let
\begin{align}
{\bar\theta} = \sqrt{\frac{|x_0|}{\pi}}\frac{\ell_p}{L^2}  .
\label{b3}
\end{align}
It is easy to check that if $\theta<-{\bar\theta}$, the right-hand
side of (\ref{a2}) is always negative, and the expansion will 
necessarily be forced down to $-\infty$.  If $\theta>{\bar\theta}$, 
the right-hand side is also negative, and the expansion will be forced 
down to $\bar\theta$.   For $-{\bar\theta}<\theta<{\bar\theta}$, 
on the other hand, the sign of the right-hand side is not fixed, and 
the expansion will vary as the vacuum fluctuates.

Now, most vacuum fluctuations are negative, and detailed
simulations in two dimensions show that most of the time they 
drive the expansion to remain near its ``maximum'' $\bar\theta$.
(Note that this is an extremely small value.)  To force runaway focusing, 
we need a positive fluctuation large enough that in a time $L$,
\begin{align}
\Delta\theta = \frac{d\theta}{d\lambda} L < -2{\bar\theta}  .
\label{b4}
\end{align}
From the Raychaudhuri equation, neglecting the effects of shear
and assuming $L\gg\ell_p$, this requires that
\begin{align}
x > 4\sqrt{\pi|x_0|}\,\frac{L}{\ell_p} = x_1  .
\label{b5}
\end{align}
This is a fluctuation large enough that a sphere of radius 
$L$ contains a mass
\begin{align}
\frac{m\,}{m_p} \gtrsim \frac{1}{3}\sqrt{\frac{|x_0|}{\pi}} \sim .13  .
\label{b6}
\end{align}
By (\ref{bx}), though, such a fluctuation is still weak, in the 
sense that
\begin{align}
\frac{Gm}{L} \sim .13 \frac{\ell_p}{L} \ll 1 \quad\hbox{for $L\gg\ell_p$}  .
\label{b7}
\end{align}

While this method of estimation may seem rather crude, it has been
shown to be extremely accurate in two dimensions \cite{CMP}.  Since 
fluctuations as large as (\ref{b5}) certainly can occur, runaway focusing is
possible.  Two questions remain, though: what is the physically relevant 
scale $L$, and, given such a scale, what is the probability of a positive 
energy fluctuation as large as (\ref{b5})?

\section{Probabilities}

We begin with the more technical question: what is the probability of
a fluctuation satisfying (\ref{b5})?  As long as $L/\ell_p$ is
reasonably large, we will be in the tail, with a probability distribution
(\ref{b1}).  Thus
\begin{align}
\mathit{Prob}(x>x_1) \sim c_0\int_{x_1}^\infty (x-x_0)^{-2} e^{-a(x-x_0)^{1/3}} dx
   = 3c_0 a^3\Gamma[-3,a(x_1-x_0)^{1/3}]
\label{c1}
\end{align}
where $\Gamma$ is an incomplete gamma function.  For large arguments,
$\Gamma(k,u)\sim u^{k-1}e^{-u}$, so
\begin{align}
\mathit{Prob}(x>x_1) \sim \left(L/\ell_p\right)^{-4/3}e^{-\alpha(L/\ell_p)^{1/3}}
\label{c2}
\end{align}
where $\alpha$ is a number of order unity whose exact value will depend
on such details as the choice of smearing function.   Thus while the probability 
of a large fluctuation drops quickly as $L$ increases, it falls off considerably
more slowly than one might first suppose.

Given such a fluctuation, we can next investigate the subsequent behavior
of our pencil of light: at what distance does it focus?  For this, we return to the
Raychaudhuri equation (\ref{a2}).  As a first approximation, which I will
justify below, let us neglect any further fluctuations of the stress-energy
tensor, as well as any contributions from the shear.  Then the Raychaudhuri
equation has the well-known solution
\begin{align}
\frac{1}{\theta} = \frac{1}{\theta(0)} + \frac{1}{2}\lambda  ,
\label{c3}
\end{align}
and focusing (i.e., $\theta\rightarrow-\infty$) occurs at at a distance
\begin{align}
\lambda_f = -\frac{2}{\theta(0)} \sim \frac{2}{{\bar\theta}} 
    \sim \beta\frac{L^2}{\ell_p}
\label{c4}
\end{align}
where $\beta$ is a number of order unity.

This could be an underestimate: although negative energy fluctuations
can never drive the expansion above $-{\bar\theta}$, they might slow its 
further descent.  But this is a tiny effect.  It is not hard to check from
(\ref{a2}) that even with the inclusion of further negative energy
fluctuations,
\begin{align}
\lambda_f \le \frac{2}{\bar\theta}\,\coth^{-1}\left|\frac{\theta(0)}{\bar\theta}\right|  .
\label{c5}
\end{align}
This is very close to (\ref{c4}) as long as $\theta(0)$ is even slightly
smaller than $-{\bar\theta}$; for $\theta(0)=-1.01\,{\bar\theta}$, for 
example, it gives only a factor of $2.5$.   
 
We thus have two observationally relevant scales: $L$, the scale that
sets the size and duration of quantum fluctuations, and $\lambda_f$, the 
distance at which these fluctuations have a significant effect on the
 propagation of light.  Note that in terms of $\lambda_f$, the probability 
(\ref{c1}) becomes
\begin{align}
\mathit{Prob}(\lambda_f) 
\sim \left(\lambda_f/\ell_p\right)^{-2/3}e^{-\gamma(\lambda_f/\ell_p)^{1/6}}
\label{c6}
\end{align}
where $\gamma$ is of order unity.  Again,
the probability of a ``large'' fluctuation falls off with size much more
slowly than one might expect.

\section{Microlensing}

Suppose a vacuum fluctuation occurs along our line of sight as we
observe a star.  If  the fluctuation is at a distance of order $\lambda_f$, 
it will focus the light, causing a momentary brightening much like
that we observe in ordinary microlensing \cite{micro}.  
A fluctuation of characteristic size $L$, focusing over a distance 
$\lambda_f$, will subtend a solid angle $\Omega_{\hbox{\tiny\it vac}}%
\sim (L/\lambda_f)^2 \sim (\ell_p/L)^2$.  A source of radius $r$ at
distance $D$ will subtend an angle $\Omega_{\hbox{\tiny\it source}}%
\sim (r/D)^2$.  For the fluctuation to focus a significant portion of
the light, $\Omega_{\hbox{\tiny\it vac}}$ should not be much smaller
than $\Omega_{\hbox{\tiny\it source}}$, so we must require 
\begin{align}
L \lesssim \frac{D}{r}\ell_p , \quad 
\lambda_f\lesssim \left(\frac{D}{r}\right)^2\ell_p  .
\label{d1}
\end{align}

As an upper limit, we may be able to observe white dwarfs in the Large 
Magellanic Cloud \cite{Elson}, which would give an optimistic estimate 
of $D/r\sim\,10^{14}$, or
\begin{align}
L\sim 10^{-21}\,\mathrm{m}  .
\label{d2}
\end{align}
This would correspond to a focusing length of $ \lambda_f\sim 10^{-7}\,\mathrm{m}$;
that is, lensing would require a fluctuation occurring within less than a 
micrometer of the telescope mirror.  Moreover, although the probability
(\ref{c2}) falls off more slowly than naive expectations, for a fluctuation
of this size it is about $10^{-20000}$.  This is not a promising
setting for observation.

We could, of course, increase the probability by making $L$ smaller.
This would also enlarge the solid angle $\Omega_{\hbox{\tiny\it vac}}$, 
boosting the number of potential sources.  For $L\sim\,5\times10^5\,\ell_p$, 
for instance, the probability of a fluctuation becomes about $10^{-42}$.  
While this is still a very small number, it is the probability for a fluctuation 
of size $L$ in a time $L/c\sim 5\times10^{5}\, t_p \sim 10^{-38}\,\mathrm{s}$, 
corresponding to a rate of several events per day.

Unfortunately, though, while such fluctuations are large compared
to the Planck length, they are still tiny compared to a typical
wavelength of light.  The geometric optics approximation implicitly 
assumed in the Raychaudhuri equation fails for scales that are small
compared to the wavelength, and there is no reason to suppose that 
such fluctuations would have the same focusing effects.  But even the 
scale $L\sim 10^{-21}\,\mathrm{m}$ of eqn.\ (\ref{d2}) corresponds 
to a photon with an energy on the order of $100\,\mathrm{TeV}$.

What is ultimately fatal to this idea, though, is the time scale of the 
fluctuations themselves.  The ``best case'' (\ref{d2}) is a fluctuation that 
lasts $10^{-29}\,\mathrm{s}$.  The case $L\sim \,5\times10^5\,\ell_p$%
---about the largest value that gives a reasonable event rate---%
corresponds to a fluctuation lasting only $10^{-38}\,\mathrm{s}$.
The prospects of observing a microlensing event of such
a short duration seem dim indeed.

One possibility remains, though: the effects of vacuum fluctuations 
could accumulate over long distances.  For photons with energies 
less than about $10^{-9}\,E_p$---even allowing for $\mathcal{O}(10)$
corrections in the exponent in (\ref{c2})---the probability of a ``large'' 
fluctuation along a photon trajectory is nearly zero even at cosmological 
distances, so runaway focusing is  not expected.  Blurring  of images 
and variations in luminosity may still occur, however.  Borgman and 
Ford have studied this process for small fluctuations \cite{Borgman}, 
and find that while the effects do accumulate, the variations in the 
expansion are of order $\ell_p{}^2/L^3$, again too small to see.  

But one loophole may still be present.  Borgman and Ford considered 
an approximation in which the nonlinearity of the Raychaudhuri
equation could be neglected.  But we know from \cite{CMP} that near
the Planck scale, negative energy fluctuations can quickly
drive the expansion to the value $\bar\theta$ of eqn.\ (\ref{b3}),
where the nonlinearities become important.  For $L\gg\ell_p$
such a process would be slower, but it could still be important.

To obtain a rough estimate, consider a random walk induced
by fluctuations of the typical size $\Delta(GT)\sim \pm\ell_p{}^2/L^4$,
each lasting a characteristic time $L$.  
After $N$ steps, the average expansion will be
\begin{align}
\langle\theta\rangle \sim \sqrt{N}\, \frac{\ell_p{}^2}{L^3}  ,
\label{d3}
\end{align}
which will be of order $\bar\theta$ when $N\sim (L/\ell_p)^2$,
that is, at a distance 
\begin{align}
d \sim \frac{L^3}{\ell_p{}^2}  .
\label{d4}
\end{align}
For a cosmological source at $d\sim50\,\mathrm{Mpc}$, the 
nonlinear regime could be reached by fluctuations as large as 
$L\sim 10^{-15}\,\mathrm{m}$.  This is still too small to see, but 
no longer quite as outrageously so---it corresponds  roughly to 
the wavelength of a $1\,\mathrm{GeV}$ photon.

This is still too crude an estimate, for several reasons.  First,
the stochastic treatment of quantum fluctuations neglects
correlations.  Second, as stressed above, the real ``random 
walk'' is highly biased---step sizes and probabilities can be read
from figure 1, and are not at all uniform.  Third, the electromagnetic
field does not give the only contribution to vacuum fluctuations;
at the scales we are considering, even QCD fluctuations may
be important.  The first of these considerations will slow the 
diffusion process, requiring longer times to reach the nonlinear 
regime, but the second and third will almost certainly shorten 
the required time.  If, in fact, the nonlinear regime can be reached, 
the effect of vacuum fluctuations can be much larger than the 
estimate of \cite{Borgman}; it will require future work to see 
whether this is possible.

\section{Conclusion}

Quantum fluctuations of vacuum energy are tiny, and one would not
ordinarily expect them to have a detectable effect on the propagation
of light.  But their distribution is highly non-Gaussian, and exceptionally
large fluctuations occur much more often than one would naively
estimate.  The estimates of the preceding section indicate that even 
averaged over a sphere with a volume $10^{17}$ times the Planck 
volume, ``large'' fluctuations may occur a few times a day.  Moreover,
the Raychaudhuri equation is nonlinear, amplifying the effect of such
large fluctuations.

But the Planck length is \emph{very} small.  In the end, the
non-Gaussian and nonlinear effects seem to require energies too
high and times too short for us to detect.  By adding more 
fields, one might shift exponents by factors of order a few, but even 
this is not likely to help.  It remains possible, though, that the
cumulative effects of fluctuations may be observable, not in
microlensing but in fluctuations and ``blurring'' of images.
It has also been suggested in \cite{Afshordi}, in a rather 
different context, that vacuum fluctuations might be visible in 
the integrated Sachs-Wolfe effect.  A further investigation 
in the present context could be of interest.

Finally, these results might place new restrictions on ``TeV-scale 
gravity'' \cite{Arkani}.  For a Planck mass of, say, $10\,\mathrm{TeV}$, 
the energies for runaway focusing discussed in the preceding 
section are reduced to about  $20\,\mathrm{MeV}$.  For the 
cumulative effects described at the end of that section, the changes
are even more dramatic: for a smearing length $L\sim700\,\mathrm{nm}$,
comparable to the wavelength of visible light, the nonlinear regime
could be reached in as little as about $100\,\mathrm{kpc}$.  A more 
thorough analysis is again necessary, but this could give a significant 
new limit on TeV-scale gravity models.

\vspace{1.5ex}
\begin{flushleft}
\large\bf Acknowledgments
\end{flushleft}

This work was supported in part by Department of Energy grant
DE-FG02-91ER40674.

\end{document}